\documentclass[10pt,preprint]{aastex}

\slugcomment{Accepted for publication in The Astronomical Journal}

\shorttitle{Spitzer Observations of Var~Her~04}

\shortauthors{Ciardi et al.}

\begin{document}

\title{
Spitzer Observations of Var~Her~04:\\
Possible Detection of Dust Formation in a Super-Outbursting TOAD}

\author{David R. Ciardi}
\affil{Michelson Science Center/Caltech\\
 770 South Wilson Avenue, M/S 100-22,
Pasadena, CA 91125}
\email{ciardi@ipac.caltech.edu}

\author{Stefanie Wachter, D. W. Hoard}
\affil{Spitzer Science Center/Caltech\\
1200 East California Avenue, M/S 220-6, Pasadena, CA 91125}

\author{Steve B. Howell}
\affil{WIYN Observatory \& NOAO\\
P.O. Box 26732, 950 N. Cherry Ave. Tucson, AZ 85719}

\and

\author{Gerard T. van~Belle}
\affil{Michelson Science Center/Caltech\\
 770 South Wilson Avenue, M/S 100-22,
Pasadena, CA 91125}

\begin{abstract}

We present four MIPS (24 \micron) and two IRAC (3.6, 4.5, 5.8, and 8.0 \micron)
Spitzer observations of the newly discovered Tremendous Outburst Amplitude
Dwarf nova (TOAD) Var~Her~04 during decline from super-outburst. The four MIPS
observations span 271 days and the two IRAC observations span 211 days. Along
the line-of-sight to Var~Her~04, there is a foreground M-star within 1\arcsec\
of the variable; as a result, all of the Spitzer photometry presented in this
paper is a blend of the foreground M-star and Var~Her~04. We estimate the
quiescent level of the TOAD to be $\Delta V=4-5$ magnitudes below that of the
M-star. Based upon the spectral energy distribution and the 2MASS colors, we
find the M-star to be an M3.5V dwarf at a distance of 80-130 pc. Based upon its
outburst amplitude and quiescent apparent magnitude, we estimate the distance
to Var~Her~04 to be $200-400$ pc, suggesting that the line-of-sight foreground
star is physically unrelated to the cataclysmic variable.  All of the Spitzer
photometry is consistent with the photospheric emission of the line-of-sight
M3.5V star, except for one 24 \micron\ observation obtained after the variable
re-brightened.  This 24 \micron\ flux density is 75 $\mu$Jy ($4\sigma$) above
the preceding and following MIPS observations. We tentatively suggest that the
mid-infrared brightening of 75 $\mu$Jy may be associated with a dust formation
event in the super-outburst ejecta. Assuming a dust temperature of $100-400$ K,
we have estimated the amount of dust required. We find $10^{-13}-10^{-11}$
M$_\odot$ of dust is needed, consistent with amounts of mass ejection in TOADs
expected during super-outburst, and possibly making TOADs important
contributors to the recycling of the interstellar medium.

\end{abstract}


\section{Introduction}

Tremendous Outburst Amplitude Dwarf novae (TOADs) consist of a white dwarf
primary star and an extremely low-mass main sequence or brown dwarf secondary
star \citep[$M \lesssim 0.1\ M_\odot$; e.g.,][]{crd98}. With typical orbital
periods of a few hours or less, the secondary star is tidally locked to the
white dwarf, and overfills its Roche lobe. Material from the secondary star
is transferred via the inner Lagrange point to an accretion disk surrounding
the white dwarf. As material builds in the accretion disk, instabilities in
the accretion disk give rise to rare ``super-outbursts,''  which, relative to
the pre-outburst quiescent luminosity, are a hundred times more luminous than
outbursts from normal dwarf novae such as U Gem and SS Cyg
\citep[e.g.,][]{harrison04}. These outbursts are not thermo-nuclear, as is
the case for classical novae.  Rather, viscous heating causes the disk
temperature to rise until ionization of hydrogen occurs, causing a pressure
wave to propagate inward, pushing material onto the surface of the white
dwarf \citep{hsc95}. The release of gravitational energy powers the
super-outburst luminosity, with dramatic brightening of 6 or more magnitudes.
Super-outbursts for TOADs occur very infrequently -- on timescales of
multiple decades.

During the super-outburst, TOADs have winds \citep[$>5000$ km/s;][]{howell95}
which are strong enough to expel material from the binary system much like
classical novae \citep{long03}, and \emph{unlike} other types of dwarf novae
which rarely display outflows that exceed (or even approach) escape velocity
during outburst (e.g., U Gem, SS Cyg). The outburst declines in TOADs
typically last a few hundred days and often show a characteristic $2-3$
magnitude dip in the optical near 50 days \citep{richter92}, much like what
is seen in the lightcurves of slow novae near 100 days
\citep[e.g.,][]{gehrz98, evans03, evans05}. In slow novae, the optical dip
can be $8 - 10$ magnitudes deep, and is attributed to the formation of an
optically thick dust shell in the ejecta \citep{sar94, evans03, evans05}.

In TOADs, the optical dip is suggested to be a dramatic drop in the mass
transfer rate from the secondary star on to the outbursting accretion disk,
or a cooling wave propagating through the accretion disk, causing a cessation
of the outburst. Approximately 10 days later, the dwarf nova re-brightens
\citep{patterson02}, attributed to a continuation of the outburst, but at
lower luminosity \citep{richter92}. However, the evidence for the exact
mechanism is sparse.

Given that TOADs have winds strong enough to eject material from the system
like classical novae, an alternative explanation of the optical dip might be
the formation of dust within the super-outburst ejecta. The V band (0.55
$\mu$m) light curve for the 1995 super-outburst of the TOAD AL Comae
Berenices has an optical dip that is $2.5-3$ magnitudes deep
\citep{howell96}, while the corresponding I band (0.9 $\mu$m) dip is only $2
- 2.5$ magnitudes deep \citep{po95} -- consistent with the 0.5 magnitude
difference between the dust extinction at 0.55 $\mu$m and 0.9 $\mu$m
\citep{mathis90}.  However, no infrared observations to detect the formation
of dust during super-outburst have been attempted - until now.

On 2004 June 16, a previously unknown star in Hercules went into outburst
\citep{iau8363, price04}. Photometric observations revealed the presence of
superhumps providing a good indication that this was a super-outburst and an
estimate for the orbital period. Given the large outburst amplitude and the
very short orbital period ($P=81.8$ minutes), Var~Her~04 was identified as a
TOAD. For a given TOAD, super-outbursts are rare (timescales of decades); we
took advantage of this timely opportunity to observe the outburst with the
Spitzer Space Telescope. We obtained two observations with IRAC (3.0, 4.5,
5.6, 8.0 \micron) and four observations with MIPS (24 \micron) to investigate
for whether dust is formed in the ejecta of TOAD super-outbursts and, hence,
possibly explain the observed dip in the optical light curves of TOADs whose
origin is a mystery.

\section{Spitzer Observations}

Four observations were obtained with the 24 \micron\ channel on the MIPS
instrument \citep{rieke04}, and two observations were obtained with all four
channels (3.6, 4.5, 5.8, and 8.0 \micron) with the IRAC instrument
\citep{fazio04}.  The first three MIPS observations (hereafter, referred to
as MIPS-1, MIPS-2, MIPS-3) were each separated by approximately 30 days
beginning on 21 Aug 2004, 67 days after the peak of the outburst.  The final
MIPS observation (MIPS-4) was obtained in May 2005 almost a full year past
the outburst peak. The first IRAC  observation (IRAC-1) was obtained between
between the second and third MIPS observation, and the second IRAC
observation (IRAC-1) was obtained near the time of MIPS-4. Table
\ref{dates-tab} details the dates and number of days past outburst peak for
each of the Spitzer observations.

\begin{figure}[ht]

    \includegraphics[angle=90,scale=0.7,keepaspectratio=true]{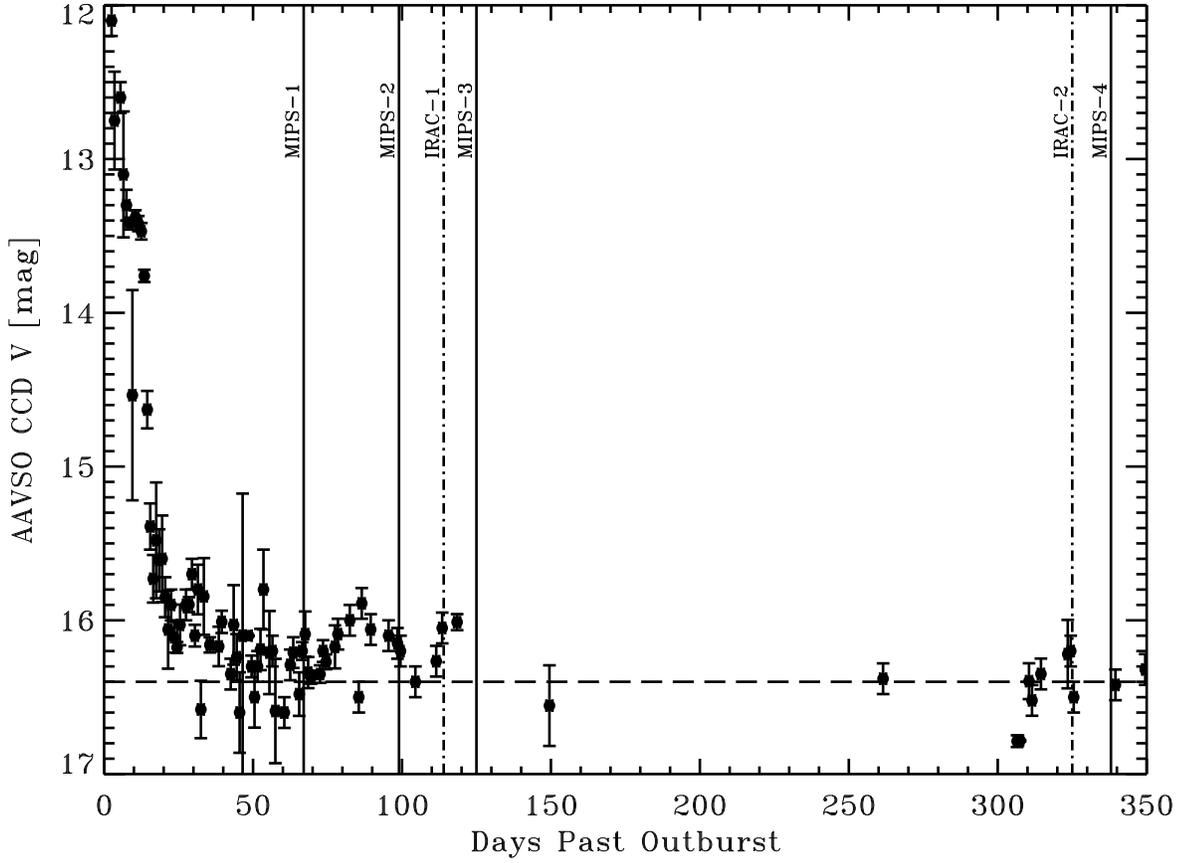}

    \figcaption{CCD visual magnitude lightcurve for Var~Her~04. Data are from
    the AAVSO and contain only CCD+V-filter validated observations.  Data have
    been re-binned in 1-day wide bins. Error bars represent dispersion of the
    observations within that bin.  The mean ``quiescent'' magnitude of V=16.5
    mag, as determined from the points at $t > 250$ days, is marked by the
    horizontal dashed line.  This is {\em not} the quiescent level of the
    outbursting Var~Her~04, but is the visual magnitude of the unrelated,
    line-of-sight foreground M-star. The times of the four
    MIPS and two IRAC observations are marked by the vertical lines.
    \label{aavso-fig} }

\end{figure}

The operational mode of the Spitzer spacecraft is to observe in
advance-scheduled instrument campaigns; i.e., a single instrument for a
dedicated period of time.  As a result, while our requests of monthly spacing
between the observations were generally met, we had no control over exactly
when those observations were made. Figure \ref{aavso-fig} displays a
validated V-band lightcurve obtained from the  American Association of
Variable Star Observers (AAVSO). The times of the MIPS and IRAC observations
are marked to indicate when in the outburst the observations were made.  As
discussed in \S3.1, along the line-of-sight to Var~Her~04, there is an M-star
within 1\arcsec\ of the variable.  None of the photometric measurements
presented in this paper (including the AAVSO lightcurve) spatially resolve
the foreground M-star from Var~Her~04.

After decline from the primary outburst, there are two minor re-brightening
($\Delta V \approx 0.5$ mag) events peaking at $t=85$ days and $t\approx 120$
days (Fig.~\ref{aavso-fig}).  The full amplitude of these two re-brightenings
is uncertain as the photometry is dominated by the M-star when Var~Her~04 is
faint.  The MIPS-1 observation ($t=67$ days) was made just prior to the first
re-brightening. The MIPS-2 observation ($t=99$ days) was obtained during
decline from the first re-brightening. The IRAC-1 ($t=114$ days) observation
was made 11 days prior to the second re-brightening peak, and MIPS-3 ($t=125$
days), and the MIPS-3 observation was during decline from the second minor
re-brightening. There are no AAVSO validated CCD+V data corresponding to the
time of MIPS-3, but interpolation suggests that the lightcurve may have been on
a slow decline between $t=125$ and $t=150$ days. The final Spitzer observations
(IRAC-2 and MIPS-4) were obtained well after the outburst event had ended, and
represent the photometric levels of the foreground M-star (see \S 3.1 below).

The IRAC data were processed with the S13.2.0 pipelines.  The MIPS-1, MIPS-2,
and MIPS-3 data were processed with the S10.5.0 pipelines, and the MIPS-4
data were processed with S12.0.2 pipeline.  The Basic Calibrated Data (BCD)
for each Astronomical Observation Request (AOR) were subsequently mosaiced
using MOPEX \citep{mm05} to create one single deep image for each of the IRAC
and MIPS visits.  The total on-source integration time was 570 seconds for
each IRAC observation, consisting of 19 dithered 30 second frames.  During
each MIPS 24 \micron\ observation, Var~Her~04 was observed for a total of
3000 seconds, consisting of 20 photometry cycles with an exposure time of 10
seconds per frame. Source identification and aperture photometry was
performed using an IDL version of DAOPHOT.  Because the MIPS-4 observation
was reduced with a slightly different pipeline sequence, only differential
photometry was performed on all of the MIPS frames to allow for direct
comparison between the visits.

\begin{figure}[ht]

    \includegraphics[angle=0,scale=0.7,keepaspectratio=true]{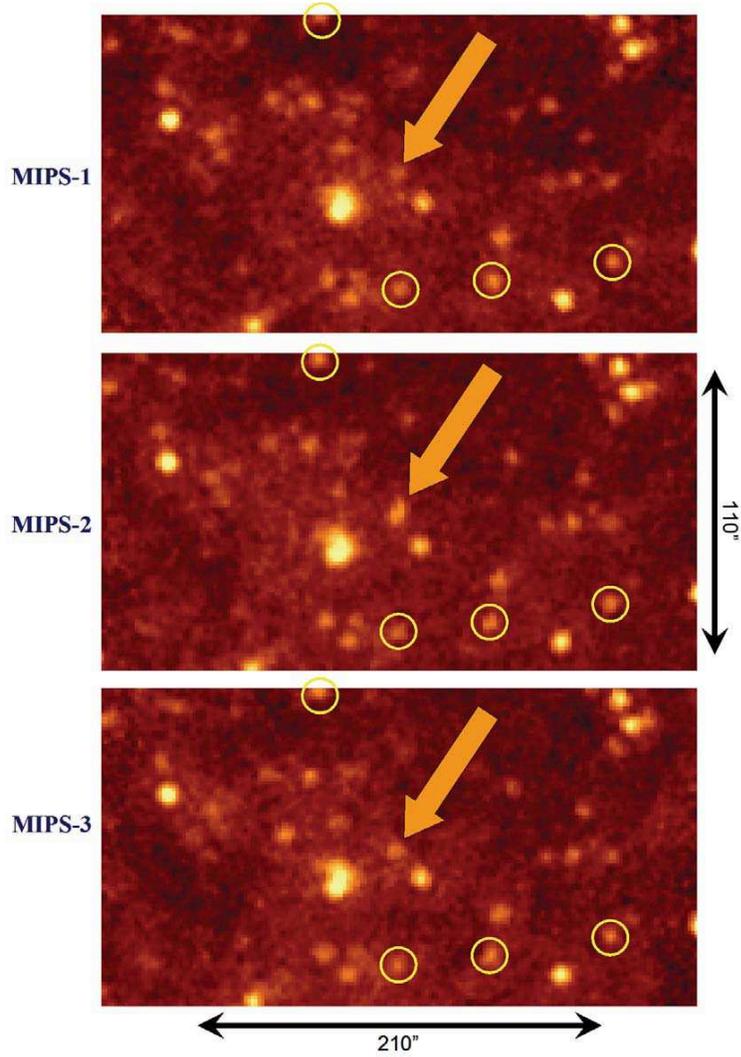}

    \figcaption{24 \micron\ ``cut-out'' images for the MIPS-1, MIPS-2, and MIPS-3
    observations. The images, oriented in equatorial coordinates with north up
    and east to the left, span $210\arcsec \times 110\arcsec$ with a pixel
    scale of $2.45\arcsec\ {\rm pixel}^{-1}$.  The arrow highlights the position
    of Var~Her~04, emphasizing the appearance of the CV in MIPS-2, and its
    absence in MIPS-1 and MIPS-3.  MIPS-4, not shown, appears identical to
    MIPS-3. The circled stars are the four stars used for comparison in the
    ensemble differential photometry. \label{mips-fig} }

\end{figure}

Figure \ref{mips-fig} displays ``cut-outs'' of the MIPS-1, MIPS-2, and MIPS-3
images centered on Var~Her~04.\footnote{MIPS-4 appears nearly identical to
MIPS-1 and MIPS-3, and is not shown.} In the figure, the arrows mark the
position of the Var~Her~04, and the circles indicate the stars used for
photometric comparison in the ensemble differential photometry.  The
equatorial coordinates for each of the comparison stars and the variable are
given in Table \ref{coord-tab}.  These comparison stars were chosen because
they were within a few arcminutes of Var~Her~04, were of comparable
brightness to Var~Her~04 at 24 \micron, and were apparently photometrically
stable at 24 \micron.  The comparison were stars were weighted and combined
into an ensemble comparison star \citep[e.g.,][]{everett02} and used to
perform standard differential photometry on Var~Her~04.

\section{Discussion}

\subsection{Distance to Var~Her~04 and the Foreground Star}

Var~Her~04, a super-outbursting TOAD, has an orbital period ($P=81.8$ minutes)
and a mass ratio ($q=0.072$) very similar to that of WZ Sge \citep{price04}.
During its 2001 outburst, WZ Sge had a peak outburst magnitude near $V=8$ mag
\citep{patterson02}, while Var~Her~04 had a peak outburst magnitude of $V=12$
mag (see Figure \ref{aavso-fig}).  TOADs have an average outburst amplitude of
$\Delta V \approx 7.5 \pm 0.8$ mag \citep{hsc95}. If we assume that WZ Sge and
Var~Her~04 have the same intrinsic outburst amplitude \citep[as most TOADs do;
e.g.,][]{howell95}, we can scale the properties of WZ Sge \citep[$d=43.5\pm0.3$
pc][]{harrison04} to Var~Her~04. At $\sim 20-40$ times fainter, the implied
distance of Var~Her~04 is $\sim 4.5-9.0$ times further away than WZ Sge for an
estimated distance of $d\approx200 - 400$ pc.

There is a foreground star located $\approx 1\arcsec$ southeast of
Var~Her~04.  Because of the line-of-sight proximity of the foreground star,
all of the photometry presented in this work is an unresolved blend of
Var~Her~04 and the foreground star. To understand the contribution of the
foreground star to our photometry, we retrieved the near-infrared photometry
at the position of the foreground star (2MASS J18392619+2604087) from the
Two-Micron All Sky Survey (2MASS; \citealt{skrutskie06}) archive.  The 2MASS
star has near-infrared magnitudes of $J=13.399 \pm 0.036$, $H=12.853 \pm
0.043$, and $K_{\rm s}=12.542 \pm 0.033$ mag. The quiescent near-infrared
brightness for the resolved variable Var~Her~04 is $J\approx17$ mag
\citep{price04}. Thus, while the 2MASS magnitudes are an unresolved blend
between the foreground star and Var~Her~04, the variable contributes less
than 3\% to the total near-infrared brightness, and the magnitudes, as
measured by 2MASS, are dominated by the foreground star.

\begin{figure}[ht]

    \includegraphics[angle=0,scale=0.7,keepaspectratio=true]{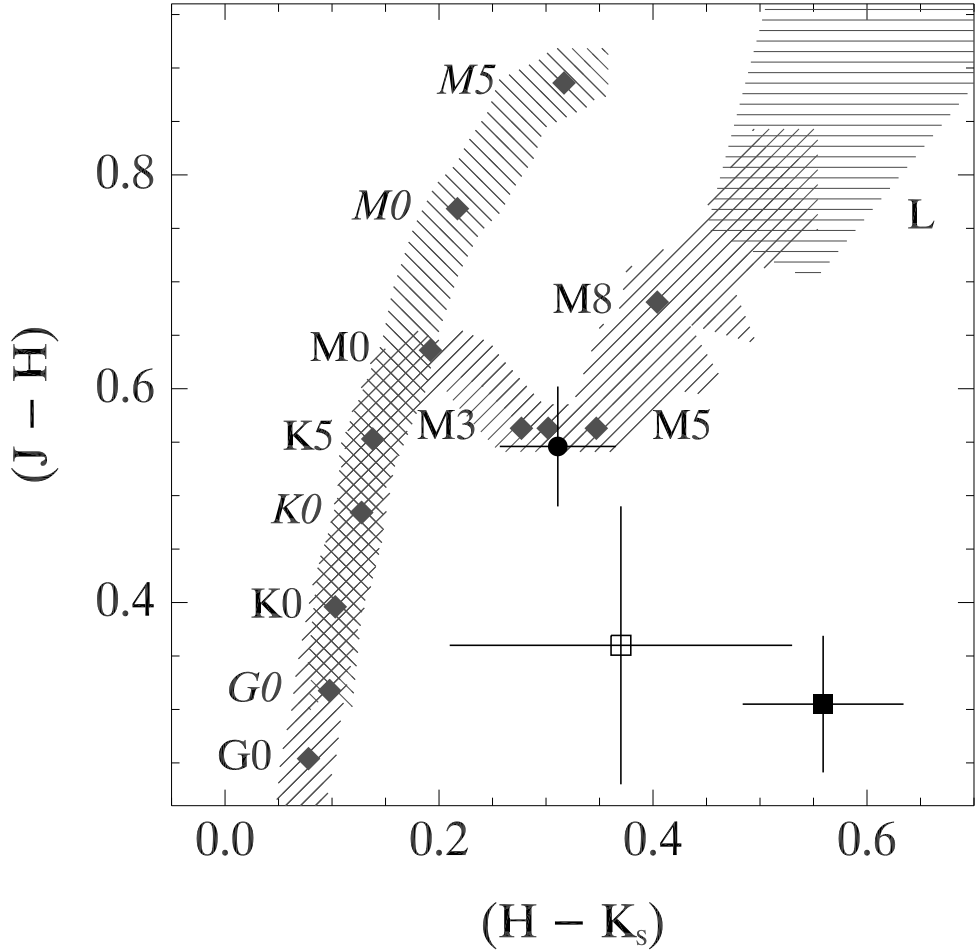}

    \figcaption{2MASS color-color diagram for dwarf and giant stars. The dwarf
    stars occupy the lower branch and separate from the giant stars at a
    spectral type of M0.  The 2MASS color photometry of the foreground star is
    marked by the filled circle.  The 2MASS color of the TOAD WZ Sge (in
    quiescence) is marked by the filled square.  The position of Var~Her~04 (in
    quiescence) is also marked (open square), showing that it has similar
    colors to WZ Sge.  Note that the JHK photometry for the quiescent
    Var~Her~04 is from \citet{price04} and is not in the native 2MASS
    magnitudes.  Diamonds mark the fiducial spectral types. \label{2mass-fig} }

\end{figure}

We have plotted the 2MASS colors in the $JHK_s$ color-color diagram shown in
Figure \ref{2mass-fig}. Because of the divergence of the near-infrared colors
of giants and main sequence stars for spectral types later than about K5, the
2MASS photometry allows us to identify unambiguously the foreground star as a
main sequence object with spectral type M4$\pm$0.5.  The near-infrared colors
for the short-orbital-period dwarf nova WZ Sge could not be mistaken for
those of a normal, low mass main sequence star (see Figure \ref{2mass-fig}).
The near-infrared colors of WZ Sge are quite similar to those reported by
\citet{price04} for the resolved quiescent variable (\ref{2mass-fig}).

Comparison of the 2MASS J-band magnitude of the foreground star with the
absolute magnitudes expected for M3 -- M4 dwarf stars \citep[$M_{\rm
J}\approx 7.8 - 8.9$,][]{hawley02}, we find a distance range of $d\approx80 -
130$ pc for the foreground star which is substantially closer than
Var~Her~04, and suggests that the foreground star is merely line-of-sight and
physically unrelated to Var~Her~04.

Scaling the quiescent brightness of WZ Sge \citep[$V=15.5$
mag,][]{patterson02}, Var~Her~04 has an expected quiescent brightness of
$V\approx 19.5$ mag, suggesting that the $V=16.5$ mag floor in Figure
\ref{aavso-fig} (marked by the horizontal dashed line) is {\em not} the
quiescent level of Var~Her~04, but rather is the visual magnitude of the
foreground M-star.

With an expected quiescent value of $V\approx19.5$ mag and an outburst
amplitude of $\Delta V\approx 7.5$ mag, Var~Her~04 is similar in brightness to
the TOAD AL Comae Berenices \citep[$V\approx20$ mag, $\Delta V\approx 8$ mag;
][]{howell96}. AL Com is at a distance of $\approx$ 250 pc \citep{shm96},
consistent with the distance estimated for Var~Her~04 from the peak outburst
brightness.

\subsection{Mid-Infrared Photometry}

\begin{figure}[ht]

    \includegraphics[angle=90,scale=0.7,keepaspectratio=true]{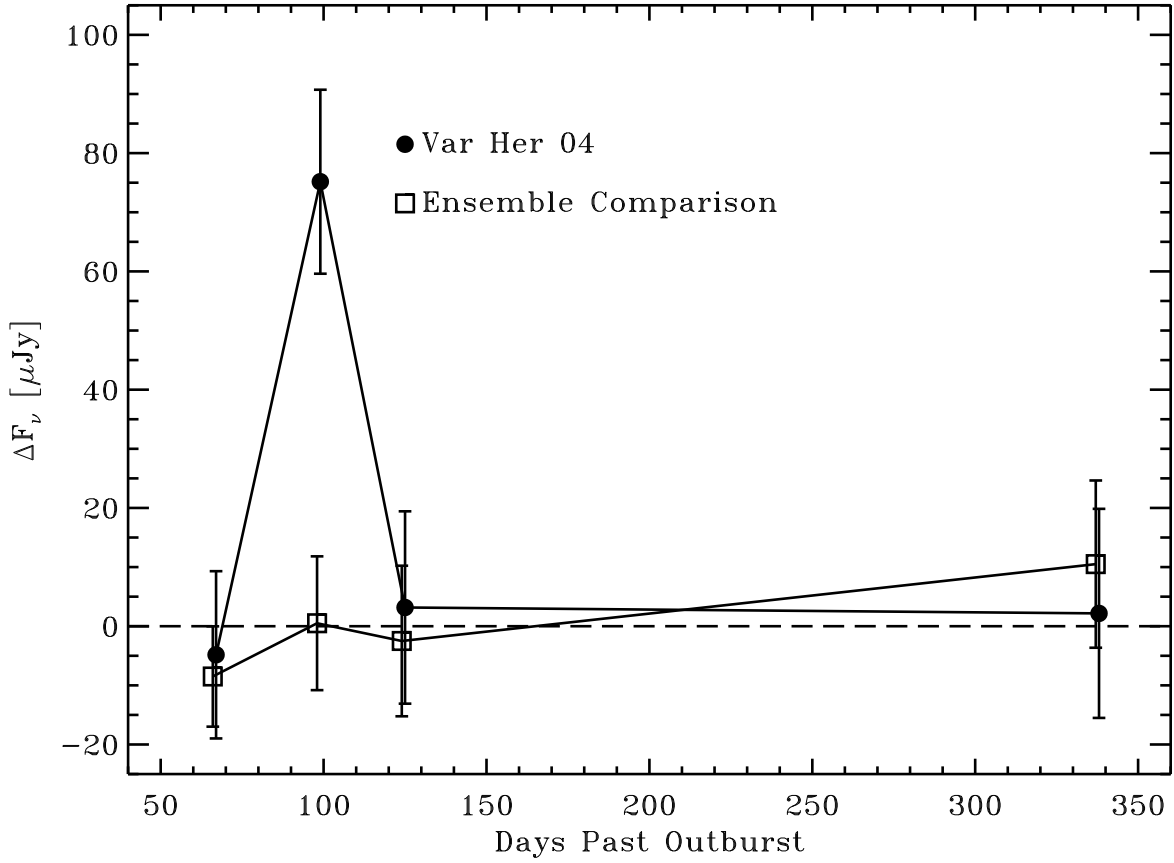}

    \figcaption{Differential 24 \micron\ photometry from the MIPS observations. The
    mean of the ensemble stars has been set to zero, as marked by
    the horizontal dashed line.  For the CV, the mean level of the 24 \micron\
    flux densities (for MIPS-1, MIPS-3, and MIPS-4) has been set to zero after
    subtraction of the ensemble flux densities. \label{mipsdiff-fig} }

\end{figure}

The differential lightcurve for the four MIPS observations (Figure
\ref{mipsdiff-fig}) clearly shows that at the time of MIPS-2 ($t=99$ days),
Var~Her~04 is 75 $\mu$Jy brighter than it is at the time of MIPS-1, MIPS-3,
and MIPS-4. The ensemble comparison does not exhibit the same brightness
change, and remains within 1$\sigma$ of the ``zero-level.''  The MIPS-2
photometric point is $\sim 4\sigma$ away from the zero-level baseline. The
brightness level increase in MIPS-2 can be seen by eye in Figure
\ref{mips-fig}.  MIPS-4 was obtained $t=338$ days after the outburst event
and sets the baseline for the M-star flux density at 24 \micron.  The
agreement of MIPS-1 and MIPS-3 with MIPS-4 implies that the two former
observations are primarily of the foreground M-star without a significant
contribution from Var~Her~04. Only the 24 \micron\ flux density at MIPS-2 is
significantly different.

For the two IRAC observations, absolute aperture photometry was performed.
The ensemble comparison stars were used to check for changes between the
IRAC-1 and IRAC-2 observations, but no significant deviations were detected.
As IRAC-2 was obtained long after the outburst event had ended, the agreement
between the IRAC-1 and IRAC-2 photometry, separated by over 200 days (see
Figure \ref{aavso-fig}), indicates that both IRAC-1 and IRAC-2 are dominated
by the foreground M dwarf and contain little (if any) contribution from the
Var~Her~04.

\begin{figure}[ht]

    \includegraphics[angle=90,scale=0.7,keepaspectratio=true]{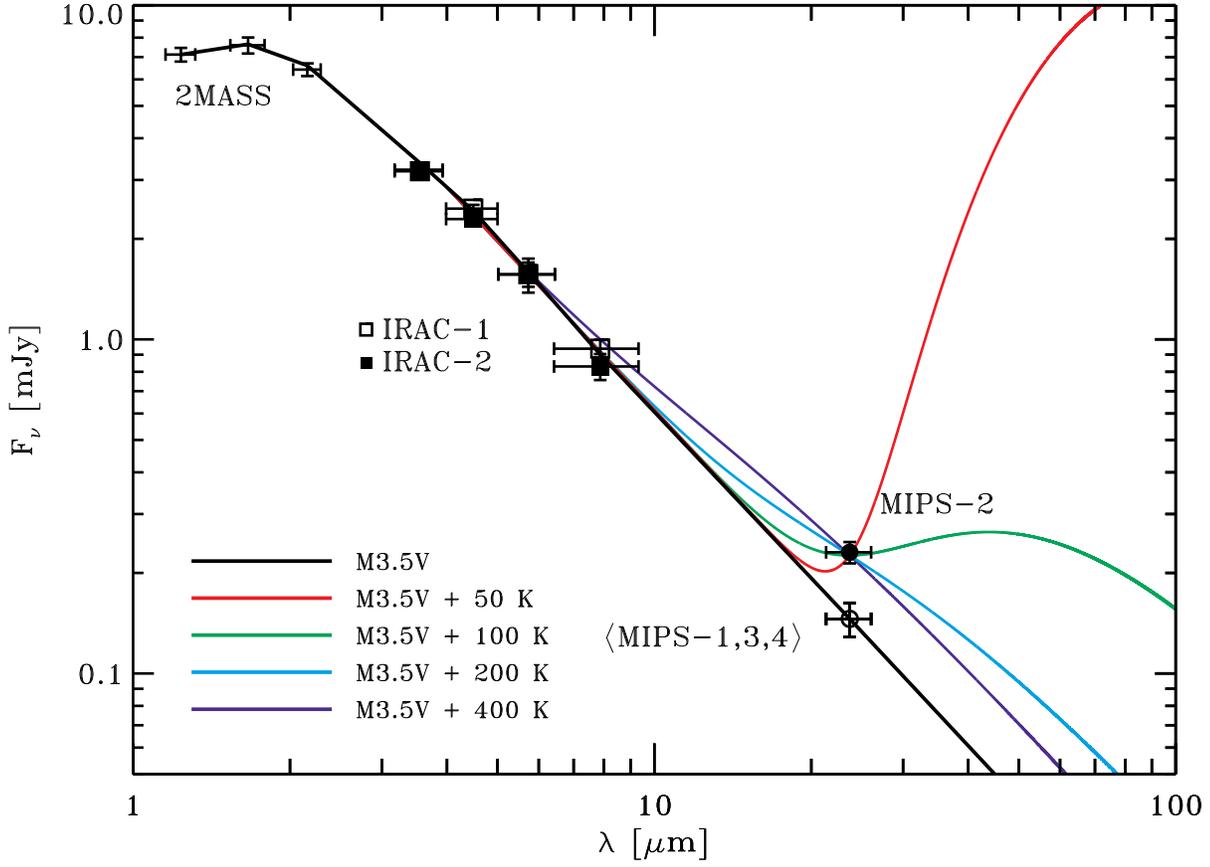}

    \figcaption{Spectral energy distribution for the unresolved photometry of
    the CV and the line-of-sight foreground M-dwarf. An M3.5V dwarf template,
    scaled to the 2MASS J-band photometric point, is shown in black. The
    ``high-state'' MIPS-2 data point is also shown, with four different
    blackbody curves forced to fit the MIPS-2 data point.\label{sed-fig} }

\end{figure}

We scaled an empirical template for the 2MASS and IRAC flux densities of an
M3.5V star \citep{patten06} to the 2MASS J-band flux density (Figure
\ref{sed-fig}).  The resulting reduced chi-square of the M3.5V template fit
to the 2MASS-IRAC data is $\chi^2_\nu \approx 1$. M3V and M4V templates were
also fitted but resulted in significantly worse chi-squares ($\chi^2_\nu
\gtrsim 10$). The goodness of the M3.5V fit for both the 2MASS and the IRAC
data indicates that the IRAC data (like the 2MASS) are dominated by the M3.5V
star, with little contribution from Var~Her~04.

The MIPS 24 \micron\ differential photometric data (Figure
\ref{mipsdiff-fig}) were scaled such that the weighted mean of the MIPS-1,
-3,and -4 data equaled the M3.5V template extrapolated to 24 \micron.  All of
the photometry from the four MIPS campaigns and the two IRAC campaigns are
reported in Table \ref{fluxes-tab}.  In the following section, we discuss the
source the 24 \micron\ brightening seen in MIPS-2.

\subsection{The 24 \micron\ Brightening}

Using Figure \ref{aavso-fig} as a reference, the optical (V) emission begins
to rise just again after the MIPS-1 observation ($t=67$ days) and after the
initial decline from the primary outburst. The cause of this re-brightening
is unknown but may be a continuation of the outburst event. At $t=85$ days,
the V emission peaks and begins to decline again with a minimum near $t=105$
days. The MIPS-2 observation at $t=99$ days is nearly at this minimum.

The MIPS-3 observation ($t=125$ days) was obtained on a day with no validated
AAVSO CCD+V data.  However, if we assume that the decline rate in V from
$t=115$ days to $t=125$ days equals the climb rate in V from $t=105$ days to
$t=115$ days, then the V magnitude at the time of MIPS-3 equals the V magnitude
at the time of MIPS-2.  In fact, we expect the V-band decline rate to be {\em
slower} than the V-band climb rate \citep[e.g., AL Com][]{howell96}. Thus, the
V-band brightness at the time of MIPS-3 is likely {\em brighter} than the
V-band brightness at the time of MIPS-2. We note here that the true rates of
rise and decline and the true depths of the minima are diluted by the presence
of the unresolved foreground M-star.

The 24 \micron\ emission associated with MIPS-3 is less than the 24 \micron\
emission at MIPS-2, even though the V magnitudes at the time of MIPS-3 is
(likely) brighter than or equal to V magnitude at the time of MIPS-2. Further,
the 24 \micron\ emission at MIPS-3 equals the 24 \micron\ emission associated
with MIPS-1 and MIPS-4, suggesting that the 24 \micron\ emission observed at
MIPS-1, -3, and -4 are dominated by the foreground M-star. Thus, the 24
\micron\ increase observed at $t=99$ days (MIPS-2) occurs, {\em not} because
Var~Her~04 simply has gotten brighter, but rather there is an additional
emission source associated with Var~Her~04.

It is conceivable that the foreground M-star could have undergone a flare event
exactly at the time of the MIPS-2 observation. However, as flare events
generally last a few hours or less \citep[e.g.,][]{rockenfeller06}, the MIPS-2
visit would have to coincide exactly with the flare.  Additionally, we
inspected the first and second halves of the MIPS-2 data cube to search for
variability within the MIPS-2 observation, as might be expected if the M-star
had undergone a flare event. We found the photometry of Var~Her~04 for the
split data to agree within $1-2\ \mu$Jy indicating there is no significant
change in brightness during the MIPS-2 observation. Consequently, we attribute
the 24 \micron\ brightening solely to the variable Var~Her~04.

We hypothesize that Var~Her~04 may have undergone a dust formation event (near
$t=85$ days) during the optical re-brightening event, in a manner very similar
to what occurs in slow-novae dust formation events \citep[e.g.,][]{evans05}.
Var~Her~04, after a steady decline since peak outburst. Var~Her~04 begins to
re-brighten ($t=65$ days).  Although the exact rate and level of the decline
and full re-brightening is lost in the confusion of the foreground star),
typical re-brightening in TOADs is $\Delta V \approx 2-3$ magnitudes
\citep[e.g., WZ Sge \& AL Com:][]{richter92, khp96}.

In our hypothesis, the re-brightening event at $t=65$ days may be sufficient to
allow dust production to occur (at $t=85$ days), and the amount of dust
produced is sufficient to obscure the optical emission from Var~Her~04, causing
the V magnitude to drop. At $t=105$ days, the dust is at its densest and the
optical emission is at a minimum.  The exact level of this dip is lost in the
glare of the foreground M-star.  It is at this conjuncture that the infrared
emission from the dust is at its peak brightness (MIPS-2).

The dust shell continues to expand. As the dust dissipates, the optical
emission from Var~Her~04 is unveiled and the V-band lightcurve re-brightens.
As the dust dissipates, the 24 \micron\ emission declines (MIPS-3) and
returns to the pre-event level (MIPS-1) which also equals the final quiescent
level (MIPS-4).   If this proposed scenario is correct, the dust forms and
dissipates in approximately $\Delta t=50$ days ($t\approx 70$ days to
$t\approx 120$ days).

Without a contemporaneous flux density at a longer wavelength (e.g., 70
\micron), we cannot constrain the dust temperature $T_d$.  We can assume,
however, a reasonable set of dust temperatures (50, 100, 200, and 400 K). In
Figure \ref{sed-fig}, we display the composite spectral energy distribution
with the 50, 100, 200, and 400 K blackbody curves overlayed (recall that the
2MASS and Spitzer photometry does not spatially resolve the foreground star
from Var~Her~04). The 50, 100, 200, and 400 K blackbody curves predict 70
\micron\ flux densities in outburst of 10, 0.2, 0.03, and 0.02 mJy,
respectively.

Without a 70 \micron\ (or longer) flux density, we cannot differentiate
between dust temperatures, but we can estimate the amount of dust required to
produce the 75 $\mu$Jy brightening at 24\micron\ for each of these dust
temperatures via
\begin{equation}
M_d = \frac{(16/3)\pi a \rho D^2}{Q_\nu B_\nu(T_d)}F_\nu
\end{equation}
where $F_\nu$ is the observed flux density at frequency $\nu$, $Q_\nu$ is the
grain emissivity at frequency $\nu$, $a$ is the grain radius, $\rho$ is the
grain mass density, $D$ is the distance to Var~Her~04, and $B_\nu$ is the
Planck function at dust temperature $T_d$. Assuming $a=0.5$ \micron, $\rho = 1$
g cm$^{-3}$, $Q_\nu = 0.1(\lambda/\micron)^{-\alpha}$, and $\alpha=0.45$
\citep[e.g.,][]{mkv06}, we estimate dust masses of $10^{-8},\ 10^{-11},\
10^{-12}\ {\rm and,}\ 10^{-13}$ M$_\odot$ for dust temperatures of 50, 100,
200, and 400 K, respectively.

For an outburst amplitude of $\Delta V = 7.5$ mag, a typical TOAD accretes
approximately $10^{-10}-10^{-9}$ M$_\odot$ on to the white dwarf surface during
a super-outburst \citep{osaki96, cannizzo01}. For a re-brightening event of
$\Delta V = 2-3$ mag, the expected amount of material ejected just from this
event may be $50-100$ times less. Even if all of this material is ejected from
the white dwarf surface, {\em and} all of that material forms dust, there is
not enough mass to explain the 24 \micron\ emission if the dust temperature is
less than $\sim 100$ K. However, if a few percent of this material is ejected
from the white dwarf surface and condenses into dust grains outside the
sublimation radius, the derived dust masses for dust temperatures of $100 -
400$ K are consistent with the anticipated accretion amounts during a
super-outburst event.

Theory predicts that there are 10$^4$ - 10$^5$ TOADs throughout the Galaxy
\citep[e.g.,][]{hns01}. If each TOAD outbursts, on average, once every 10
years, then there may be 10$^3$ - 10$^4$ outbursts per year throughout the
Galaxy.  If each outburst produces $10^{-13} - 10^{-11}$ M$_\odot$ of dust,
this corresponds to $10^{-10} - 10^{-8}$ M$_\odot$ of dust injected into the
interstellar medium per year.  On average there is one classical nova per
year in the Galaxy, and each classical nova ejects $10^{-9} - 10^{-6}$
M$_\odot$ of dust into the interstellar medium \citep{gehrz98}, perhaps
making the total population of TOADs as important as classical novae in the
recycling of the interstellar medium.

\section{Summary}

We have obtained Spitzer MIPS and IRAC observations of the newly discovered
Tremendous Outburst Amplitude Dwarf nova (TOAD) Var~Her~04 in decline from
super-outburst. Four MIPS observations at 24 \micron\ were made spanning 271
days.  In addition, two sets of IRAC observations (all four bands) spanning
211 days were also made. All of the Spitzer photometry is consistent with the
photospheric emission of a line-of-sight (but physically unrelated) M3.5V
star, except for one 24 \micron\ observation obtained after the variable
re-brightened.  We tentatively suggest that the mid-infrared brightening of
75 $\mu$Jy may be associated with a dust formation event.  We estimate that
the amount of dust required to produce such a brightening at 24 \micron\ is
only $10^{-13}-10^{-11}$ M$_\odot$, consistent with the amount of material
expected in super-outburst ejecta. Given the total population of TOADs in the
Galaxy, TOADs may be as important as classical novae in terms of their
production of dust.  A dedicated observing program to study TOADs (that are
not spatially confused with foreground stars) in super-outburst is needed to
understand more clearly the ejecta processes, and the contribution of TOADs
to the recycling of the interstellar medium.

\acknowledgments

The authors would like to thank the Director of the Spitzer Science Center
for granting our observing request through the Director's Discretionary Time.
This work has been supported, in part, by NASA through the Spitzer Science
Center and the Michelson Science Center at Caltech.  A special thank you to
Steve Schurr for his enthusiasm and red pen.

We acknowledge with thanks the variable star observations from the AAVSO
International Database contributed by observers worldwide and used in this
research.  This research has made use of the NASA/IPAC Infrared Science
Archive, which is operated by the Jet Propulsion Laboratory, California
Institute of Technology, under contract with the National Aeronautics and
Space Administration.  This publication makes use of data products from the
Two Micron All Sky Survey, which is a joint project of the University of
Massachusetts and the Infrared Processing and Analysis Center/California
Institute of Technology, funded by the National Aeronautics and Space
Administration and the National Science Foundation.

\begin{deluxetable}{cccc}
\tablecolumns{4}
\tablewidth{0pc}
\tablecaption{Dates of Spitzer Observations \label{dates-tab} }

\tablehead{ \colhead{Observation} & \colhead{UT Date} & \colhead{Days Past
Outburst} & \colhead{AOR Key}} \startdata
MIPS-1 & 21 Aug 2004 & 67 & 12000000\\
MIPS-2 & 22 Sep 2004 & 99 & 12000768\\
IRAC-1 & 07 Oct 2004 & 114 & 12001024\\
MIPS-3 & 18 Oct 2004 & 125 & 12001280\\
IRAC-2 & 06 May 2005 & 325 & 13533696\\
MIPS-4 & 19 May 2005 & 338 & 13533440\\
\enddata

\end{deluxetable}

\begin{deluxetable}{ccc}
\tablecolumns{4} \tablewidth{0pc}
\tablecaption{Coordinates \label{coord-tab}
}

\tablehead{ \colhead{Star} & \colhead{R.A. (J2000)} & \colhead{Dec.
(J2000)}\\

\colhead{} & \colhead{[hh:mm:ss]} & \colhead{[dd:mm:ss]} } \startdata
Var~Her~04   & 18:39:26 & 26:04:08\\
Comparison-1 & 18:39:20 & 26:03:39\\
Comparison-2 & 18:39:24 & 26:03:32\\
Comparison-3 & 18:39:26 & 26:03:29\\
Comparison-4 & 18:39:28 & 26:05:07\\
\enddata

\end{deluxetable}

\begin{deluxetable}{lcc}
\tablecolumns{4} \tablewidth{0pc} \tablecaption{Flux
Densities\tablenotemark{a} \label{fluxes-tab} }

\tablehead{ \colhead{Bandpass} & \colhead{Flux Density} &
\colhead{Observation}\\

\colhead{[$\mu$m]} & \colhead{[mJy]} & \colhead{ID}
} \startdata
1.235 (0.162) & $7.12\pm0.34$ & 2MASS J\\
1.662 (0.251) & $7.59\pm0.42$ & 2MASS H\\
2.159 (0.262) & $6.42\pm0.28$ & 2MASS K$_s$\\
3.550 (0.75)  & $3.19\pm0.07$ & IRAC-1\\
              & $3.22\pm0.10$ & IRAC-2\\
4.493 (1.015) & $2.46\pm0.06$ & IRAC-1\\
              & $2.29\pm0.09$ & IRAC-2\\
5.731 (1.425) & $1.57\pm0.13$ & IRAC-1\\
              & $1.56\pm0.18$ & IRAC-2\\
7.872 (2.905) & $0.94\pm0.06$ & IRAC-1\\
              & $0.83\pm0.08$ & IRAC-2\\
23.7 (4.7)    & $0.15\pm0.02$ & $\langle$MIPS-1,3,4$\rangle$\tablenotemark{b}\\
              & $0.23\pm0.02$ & MIPS-2\\

\enddata
\tablenotetext{a}{Unresolved Photometry of Var~Her~04 and Foreground
Line-of-Sight M-star.} \tablenotetext{b}{Mean of differential photometry from
MIPS-1, MIPS3 and MIPS-4 scaled to the M3.5V template.}

\end{deluxetable}

\clearpage
\clearpage

\clearpage

\clearpage

\clearpage

\end{document}